\documentclass[aps,prl,twocolumn]{revtex4}
\usepackage{graphics,epsfig,amsfonts,amssymb,amsmath}
\begin{document}

\title{On the checkerboard pattern and the autocorrelation of photoemission
data in high temperature superconductors}
\author{E. Bascones and B. Valenzuela}
\affiliation{Instituto de Ciencia de Materiales de Madrid,
CSIC. Cantoblanco. E-28049 Madrid, Spain}
\begin{abstract}
In the pseudogap state the spectrum of the autocorrelation of angle
resolved photoemission (AC-ARPES) data of
Bi$_2$Sr$_2$CaCu$_2$O$_{8+\delta}$ presents non-dispersive peaks in
momentum space which compare well with those responsible for the
checkerboard pattern found in the density of states by Scanning
Tunneling Microscopy. This similarity suggests that the checkerboard
pattern originates from peaks in the joint density of states, as the
dispersive peaks found in the superconducting state do. Here we show
that the experimental AC-ARPES spectrum can be reproduced within a
model for the pseudogap with no charge-ordering or symmetry
breaking. We predict that, because of the competition of
superconductivity and pseudogap, in the superconducting state, the
AC-ARPES data of underdoped cuprates will present both dispersive
and non-dispersive peaks and they will be better observed in
cuprates with low critical temperature. We finally argue that the
AC-ARPES data is a complementary and convenient way to measure the
arc length.
\end{abstract}
\date{\today}
\email{leni@icmm.csic.es,
belenv@icmm.csic.es}
\maketitle

In the pseudogap (PG) state of underdoped cuprates instead of a
complete Fermi surface(FS)\cite{Norman98}, just a Fermi arc around
the nodal (diagonal) direction is seen, while the antinodal region
close to $(\pi,0)$ is gapped. Raman\cite{leTacon06} and
photoemission\cite{shentwoscales06,aipes06,kaminski06} experiments
have shown that the nodal-antinodal dichotomy persists in the
superconducting (SC) state, in the form of two different energy
scales in nodal and antinodal regions. This behavior can be
explained in terms of the coexistence and competition of SC and PG
correlations below the critical temperature
($T_c$)\cite{nosotras06}. The nature of the possible competing state
remains controversial. Most of the proposals involve charge-ordering
and/or breaking of the symmetry. To date there is no accepted
evidence of such symmetry breaking. Strong support for
charge-ordering models came from the observation of the so-called
checkerboard pattern in Fourier Transform Scanning Tunneling
Spectroscopy (FT-STS)
measurements\cite{Hoffman02-1,Hoffman02-2,McElroy03,Vershinin04,Hanaguri04,McElroy05,Howald03,Levy05,Hashimoto06,Machida06}.
But this interpretation seems at odds with the data obtained from
the autocorrelation of the Angle Resolved Photoemission (ARPES)
spectra\cite{Lanzara06,Campuzano06}.
The checkerboard pattern refers to the non-dispersive peaks in the
momentum {\bf q} and energy $\omega$ dependent density of states
$n({\bf q},\omega)$ found at ${\bf q}\sim (\pm 2\pi/\lambda,0)$ and
$(0,\pm 2 \pi/\lambda)$ with $\lambda\sim 4$-$5$ in units of the
lattice spacing. Together with this modulation, weaker $3/4$
substructure at ${\bf q}\sim (\pm (2\pi)3/4,0)$ and $(0,\pm
(2\pi)3/4)$ has been
detected\cite{Hanaguri04,McElroy05,Levy05,Hashimoto06,Machida06}.

The lack of dispersion of the
checkerboard peaks differentiate them from another kind of peaks also found by
FT-STS in the SC state which disperse with binding energy.
It is generally accepted that the dispersive features
are  a consequence of quantum interference of quasiparticles by
elastic scattering\cite{dhlee03,Capriotti03,Hirschfeld}. In the quasiparticle
interference picture,
maxima in $n({\bf q},\omega)$ are expected at those momenta which
connect the states with the largest joint density of states (JDOS).
In the so-called octet model\cite{dhlee03}, in the SC state the
largest density of states at a given $\omega$ is found at the tips
of the banana-shaped constant energy contours around the nodes
(Fig.~1(a)) and $n({\bf q}, \omega)$ peaks at the wavevectors ${\bf
q}_1$,...,${\bf q}_7$
which connect such tips. The size of these banana-shape constant energy
contours changes with binding energy producing the dispersive behavior of
the peaks.
On the contrary, the origin of the checkerboard remains
controversial and highly debated.
Due to its non-dispersive nature\cite{Phillips04},
most of the
models involve inhomogeneous states and charge ordering\cite{Hanaguri04,Chen02,Tesanovic04,Anderson04,Polkovnikov02,Podolsky03,Vojta02,Li06,Ohkawa06,dhlee06,Dellanna05}, but proposals based on the JDOS picture
have been
also discussed\cite{Phillips04,Chakravarty}.

The JDOS picture can be checked\cite{Markiewicz04}  autocorrelating
ARPES data. The ${\bf q}$-space pattern measured from the
autocorrelation of ARPES (AC-ARPES) data can be directly
interpreted, as it does not require any theoretical modeling.
Neglecting the matrix element, ARPES  measures the spectral function
$A({\bf k},\omega)$. In AC-ARPES the JDOS is obtained from the
convolution
\begin{equation}
JDOS({\bf q},\omega)=\sum_k A({\bf k},\omega)A({\bf k+q},\omega).
\label{jdos}
\end{equation}
The AC-ARPES spectra of Bi$_2$Sr$_2$CaCu$_2$O$_{8+\delta}$ (Bi2212)
in the SC state\cite{Lanzara06,Campuzano06} show dispersive peaks as
those expected from the octet model.
In the PG state  AC-ARPES
data present\cite{Campuzano06}
 peaks near $(0.4\pi,0)$ with very little dispersion, in
contrast to the ones in the SC state. These non-dispersive
peaks compare well with those
responsible for the checkerboard  in FT-STS.
This similarity
point to a JDOS explanation of the checkerboard  and cast doubt on
those models involving charge ordering.
Analysis of the experimental data shows that they are associated to
vectors of the ${\bf q}^*_1$ type in Fig.~1(b), connecting
the tips of the Fermi arcs. Peaks corresponding to ${\bf q}^*_5$ in Fig.~1(b)
and structure along the diagonal are also observed.
It remains to be explained how does this
non-dispersive behavior appears in the JDOS.

In this letter we show that the experimental AC-ARPES spectrum
is well reproduced by the model recently proposed
by Yang, Rice and Zhang (YRZ) for the pseudogap\cite{YRZ}.
Neither intrinsic charge-ordering or symmetry breaking are involved
in this model or in
the explanation of the experimental results.
In agreement
with experiments we find
 peaks with very
little dispersion (referred as non-dispersive in the following) in
the PG while clearly dispersive peaks appear in the SC state. This
behavior is related to the different evolution of the constant
energy contours size with binding energy and the existence of Fermi
arcs at zero energy. Non-dispersive peaks presumably related to the
checkerboard $3/4$ substructure are also found in the PG.
Furthermore, we predict that both dispersive and non-dispersive
peaks can be present at low doping $x$ in the SC state and we relate
this result with the U-shape of the SC gap found in ARPES in
underdoped SC cuprates\cite{shentwoscales06,mesot}. The dispersive
ones are restricted to low energies, lower with underdoping. The
non-dispersive features in the SC state dominate the spectrum at
small doping and are a consequence of the persistence of PG
correlations below $T_c$ and its imprint on the spectral function
and dispersion.
\begin{figure}
\leavevmode
\includegraphics[clip,width=0.42\textwidth]{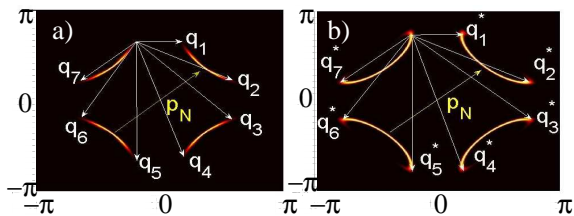}
\caption{(color online) (a) ARPES intensity around the nodes in the superconducting
  state for $\omega=-0.04$, in units of the bare nearest neighbor hopping, and
  $x=0.20$. (b) Same as in (a) in the
pseudogap state for $x=0.16$ according to the model discussed. The wavevectors
${\bf q}_i$ (${\bf q}^*_i$) of quasiparticle interference
  patterns in the octet model, as well as the intranodal momenta along the
  diagonal ${\bf P}_N$.
  }
 \label{1}
\end{figure}

In the YRZ model PG correlations are given at zero temperature by
$\Delta_R$ which does not break any symmetry because of its spin
liquid origin\cite{YRZ}. $\Delta_R$ decreases with doping $x$ and
vanishes at a topological quantum critical point $x_c$. A crucial
point in this model is the appearance of hole pockets close to $(\pm
\pi/2,\pm \pi/2)$. Due to reduced spectral weight on the outer edge
of the
 pocket, a gapless Fermi arc appears in ARPES at zero
energy\cite{YRZ,nosotras06}. At finite energy the arc structure
remains as shown in Fig.~1(b).

In the SC state the SC order parameter $\Delta_S$ is related to
$T_c$. It is assumed that PG and superconductivity coexist below
$T_c$ and $x_c$. Both $\Delta_R$ and $\Delta_S$ have d-wave symmetry
 $\Delta_\alpha({\bf k})=\Delta_\alpha(x)/2(\cos k_x-\cos k_y)$ with
 $\alpha=R,S$, but they gap the FS in a different way. At zero 
frequency the BCS
 self-energy diverges at the FS while the YRZ self-energy diverges at
 the umklapp surface $|k_x \pm k_y|=\pi$\cite{YRZ}.
Below $x_c$, we characterize the PG state by zero $\Delta_S$ and
finite $\Delta_R$. Beyond $x_c$, $\Delta_R$ vanishes and a complete
Fermi surface and BCS behavior are
 recovered.  We take
the same parameters
for $\Delta_S$, $\Delta_R$ and the band dispersion proposed in the
original paper\cite{YRZ} and used afterwards\cite{nosotras06}. In particular,
$\Delta_R(x)/2=0.3(1-x/0.2)$ and
$\Delta_S(x)/2=0.07(1-82.6(x-0.2)^2)$ with energies units
of the
bare nearest neighbor hopping $t_0 \sim 300-400 meV$.
The anomalous behavior, i.e. the
emergence of non-dispersive features, is
expected below $x_c=0.2$.
The exact expressions for the spectral function and energies of
the YRZ model have been given
elsewhere\cite{YRZ,nosotras06} and we do not repeat them here.
 To compare with experiments we calculate Eq.~(\ref{jdos})
 following the same procedure\cite{metodo} as in
 refs\cite{Lanzara06,Campuzano06}.
Experimental AC-ARPES spectra is also influenced by the anisotropic and
energy dependent
lifetime not included here.

The AC-ARPES map
in the SC state for $x_c=0.20$
is shown in Fig.~2(a) $\omega=-0.04$. It closely resemble the one
obtained from experimental data\cite{Lanzara06,Campuzano06} in the
SC state, as well as those arising from the convolution of the Green
function with itself  discussed in the context of FT-STS
experiments\cite{dhlee03,Capriotti03} but it lacks the kaleidoscopic
patterns due to umklapp\cite{Capriotti03,metodo} present in the
latest ones. The peaks corresponding to $q_i$-type terms in
Fig.~1(a) are observed. The change of momenta with binding energy of
the peaks along the bond and diagonal directions is clearly seen in
Figs.~2(d) and 2(g). The almost dispersionless peak in Fig.~2(g). is
due to nesting\cite{Capriotti03} and corresponds to ${\bf P}_N$-like
contributions, as also seen by McElroy {\it et al}\cite{Lanzara06}.
\begin{figure}
\leavevmode
\includegraphics[clip,width=0.42\textwidth]{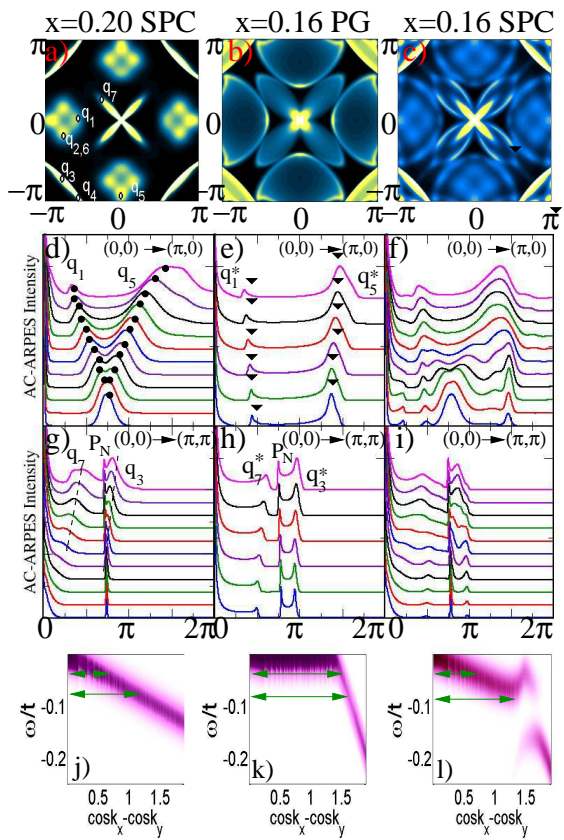}
\caption{(color online) AC-ARPES and energy spectra for
$x_c=0.20$ ($\Delta_R=0$,$\Delta_S=0.14$) in the
superconducting state (left column) and $x_c=0.16$ in the pseudogap state
($\Delta_R=0.12$, $\Delta_S=0$, middle column) and in the
superconducting state($\Delta_R=0.12$, $\Delta_S=0.12$, right column). (a)
to (c)
  show, in arbitrary units, the  maps at $\omega=-0.04$. (d) to (f), and (g) to (i) Intensity of the autocorrelated spectral
function along the
  $(0,0)-(\pi,0)$ (bond) and $(0,0)-(\pi,\pi)$ (diagonal) directions
respectively, at several energies . From bottom to
  top $\omega$=-0,to -0.10 in 0.02 intervals in the pseudogap state and in
0.01 intervals in the superconducting state, and in units of the bare
  nearest neighbor hopping. Each curve in (d) to (i) is normalized to the
value at its largest
feature other than the one at $(0,0)$ and displaced, to better show
  the peaks dispersion. The different behavior of the energy dispersion
  in the SC, PG and SC\&PG states can be seen in (j) to (l) where
  the energy spectrum along the $w=0$ maximum ARPES intensity line is plotted.
  Arrows are at $w=-0.04$ and $w=-0.08$.}
 \label{2}
\end{figure}

At first sight the ARPES intensities in the the BCS superconducting
(Fig.~1(a)) and the PG states (Fig.~1(b)) look similar. But their
AC-ARPES spectra and energy-dependence show important differences.
Contrary to what happens in the SC case the peak along the bond in
the AC-ARPES spectrum in the PG state is split in Fig.~2(e), {\it
even at zero energy}. Their momenta and energy dependence resemble
that found in the PG by Chatterjee {\it et al}\cite{Campuzano06}.
Their positions change very little with energy in strong contrast to
the dispersive behavior in Fig.~2(d). This weak dispersion could be
reduced by finite lifetimes, weaker spectral weight at the antinode
or worse experimental resolution than included here.

The checkerboard pattern is presumably related to the small momentum
peak along the bond, ${\bf q}^*_1$. We note that the momentum at
which the $3/4$ substructure has been observed\cite{Hanaguri04} is
very close to that of the ${\bf q}^*_5$ peak, and we postulate that
both features are related. Zero-energy splitting and non-dispersing
behavior also appear along the diagonal~(Fig.~2(h)). The origin of
the different behavior of peak position in the SC and PG states is
related to the dependence of constant-energy contour size with
binding energy  which can be inferred from Fig.~2(j) and 2(k) where
the corresponding energy spectrum along the $w=0$ maximum ARPES
intensity is reproduced. The length of the arrows at $w=-0.04$ and
$w=-0.08$ give an idea on the change of the constant energy contours
with binding energy. In the SC state the zero energy contour is a
single point since all the FS is gapped. The contour size increases
rapidly with binding energy. On the other hand, in the PG state the
peak splitting along the bond comes from the finite size of the
closed pocket centered around $(\pi/2\pi/2)$. The size of the
constant-energy contour barely changes with binding energy.

Dispersive behavior at low energies appears in the SC state even for
finite $\Delta_R$. Interestingly, both dispersive and non-dispersive
features can be distinguished in Fig.~2(f). Dispersive peaks, of the
type observed in Fig.~2(d), dominate at low energy but there is a
clear kink in the dispersion and the peaks in the AC-ARPES spectrum
converge to those observed in the PG state. The opening of a gap due
to superconductivity in the arcs suppresses at low energies the
non-dispersive peaks arising from the tips of the arcs but remanent
structure is visible at the corresponding momenta. The appearance of
both types of peaks is due to the coexistence of SC and PG. This
coexistence can be seen in the energy band spectrum in Fig.~2(l).
The gap in the arc at low energies is dominated by
superconductivity. At higher energies, the antinodal region is
mainly affected by the pseudogap. This energy spectrum has been
proposed\cite{nosotras06} to explain the U-shape of the SC gap found
in ARPES in underdoped SC cuprates\cite{shentwoscales06,mesot}. The
dispersive and non-dispersive features are better seen in Fig.~3(a),
where the position of the maxima is plotted. The energy at which the
change from dispersive to non-dispersive behavior happens is mainly
given by $\Delta_S({\bf k})$ at the arc tip, and depends on
$\Delta_R$, via the arc length. Overlap of dispersive and
non-dispersive peaks does not always allow to differentiate  them or
to associate  the position of the maximum in intensity to a
particular kind of peak. We note that dispersive peaks at low
energies and non dispersive ones at higher energies have been
observed in FT-STS experiments in underdoped Bi2212\cite{McElroy05}.
The AC-ARPES spectra, corresponding to $x=0.12$ and $x=0.18$, in the
SC state along the bond are shown in Figs.~3(b) and 3(c). $\Delta_S$
and $\Delta_R$ are finite in both cases. However, dispersive and
non-dispersive peaks are not as clearly identified here as they were
in Fig.~2(e). Thus the presence of only dispersive (non-dispersive)
peaks does not guarantee zero $\Delta_R$ ($\Delta_S$).

\begin{figure}
\leavevmode
\includegraphics[clip,width=0.42\textwidth]{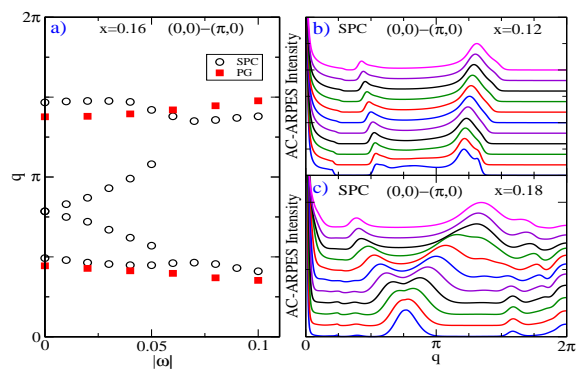}
\caption{(color online) (a) Position of the dispersing and non-dispersing
peaks found in the
  $(0,0)-(\pi,0)$ direction in the superconducting and pseudogap state at
  $x=0.16$. (b) and (c) Same as in Fig. 2(f) but for $x=0.12$
($\Delta_R=0.24$, $\Delta_S=0.07$) and $x=0.18$
($\Delta_R=0.06$, $\Delta_S=0.13$)
respectively. }
 \label{4}
\end{figure}
In general, for smaller $\frac{\Delta_S}{\Delta_R}$
the non-dispersive structure is more pronounced and for a given
doping, the range of energies at
which  dispersive features appear is
reduced with decreasing $\Delta_S$. In agreement with
recent ARPES measurements\cite{kaminski06}, we expect $\Delta_S$ to be
to some extent related to $T_c$.
Based on this argument, we predict that in the SC
state  the
peaks in the AC-ARPES spectra of low $T_c$ cuprates, like
Ca$_{2-x}$Na$_x$CuO$_2$Cl$_2$ (Na-CCOC) or
Bi$_{2}$Sr$_{2}$CuO$_{6+\delta}$ (Bi2201),
will be mostly non-dispersive, similar to the ones found in
the PG state in Bi2212\cite{Campuzano06}.
In low $T_c$ cuprates $\Delta_R$ and $\Delta_S$ are expected to differ
more and
$\frac{\Delta_S}{\Delta_R}$ to be smaller.
We note that in FT-STS experiments in the SC state,
the checkerboard pattern is better seen in
cuprates with low
$T_c$ and when the
integrated density of states lacks the coherence peaks, when it is more PG
like\cite{Hanaguri04,McElroy05,Hashimoto06,Machida06}.
In fact, to the best of our knowledge in FT-STS dispersive features
in the SC state have been seen so far only in
Bi2212\cite{Hoffman02-2,McElroy03,McElroy05}, and not in
low-$T_c$ cuprates as  Na-CCOC\cite{Hanaguri04}
or Bi$_{2}$Sr$_{1.6}$La$_{0.4}$CuO$_{6+\delta}$ (Bi2201-La)\cite{Machida06}.
The observation of dispersive peaks at very low energies in FT-STS
experiments has been maybe prevented by the small
signal-to-noise ratio at low energies.
On the other hand, ARPES measurements work well at these energies.
Confirmation of the presence
of
both dispersive and non-dispersive peaks in  AC-ARPES in the
SC state (with the dispersive ones restricted to energies of a
few meV in some cases) would confirm the existence of two energy scales and
a common origin of the peaks observed in AC-ARPES and FT-STS.

Recent ARPES experiments\cite{Kanigel06} have suggested that in the PG state
the Fermi length
is temperature dependent and vanishes at low temperatures, while the antinodal
region remains gapped up to the PG temperature.
The arc length, as measured directly from the ARPES intensity suffers from
large uncertainty.
The observation by Chatterjee {\it et al}\cite{Campuzano06} that
the nondispersive peaks found
in the AC-ARPES in the PG state arise from the tips of the Fermi arcs
and that a SC-type gap in the arcs results in dispersive features, opens a new
 way to get complementary
information on the length of the Fermi arc.
We propose that AC-ARPES experiments can be used to determine the position of
the arc tips and nodal Fermi momentum, as well as of the arc length and its
dependence with temperature and doping.
We note that as the spectra is autocorrelated the uncertainty of the position
of the arc tips is strongly reduced compared to the bare intensity. The
dependence of the arc size with energy can be also measured, providing extra
information of the physics involved in the truncation of the
Fermi surface.
We believe that low $T_c$ cuprates are the most
suitable for this experiment as the two energy scales $\Delta_R$ and $\Delta_S$
will be most different and underdoped non-SC samples, showing the
checkerboard,
are available\cite{Hanaguri04}.

In conclusion, we have shown that the non-dispersive structure found
in the autocorrelation of photoemission data in Bi2212 in the
pseudogap state can be explained without involving charge ordering
or symmetry breaking but the existence of the Fermi arcs  and a
weak binding energy dependence of the size of the constant energy
contour. We believe that both the checkerboard and its $3/4$
substructure can be explained within a joint density of states
picture. Furthermore, we predict the simultaneous appearance of both
dispersing and non-dispersing peaks in the AC-ARPES spectra in the
superconducting state of underdoped cuprates  which originates in
the U-shape of the SC gap, and will be better observed in
materials with low $T_c$. The observation of the coexistence of the
two-types of peaks would be the {\it smoking gun} which confirm the
joint density of states mechanism for the checkerboard, the
equivalence of AC-ARPES and FT-STS peaks and the coexistence of
pseudogap and superconductivity below $x_c$. We also propose to
autocorrelate ARPES data to measure the arc length and its
temperature dependence.

After the first submission of this work we have been aware of new
experimental results with confirm the prediction of coexistence of
dispersing and non-dispersing peaks in the superconducting
state\cite{Campuzano07}.

We acknowledge funding from MEC through Grant No. FIS2005-05478-C02-01 and
Ramon y
Cajal contract and from Consejer\'ia de Educaci\'on de la Comunidad de Madrid
and CSIC
through Grant No. 200550M136 and I3P contract.

\end{document}